\documentstyle[pra,aps,multicol,psfig]{revtex}
\begin{document}

\draft

\title{Critical Temperature of a Trapped Bose Gas: Mean-Field Theory 
       and Fluctuations}

\author{M. Houbiers,$^1$ H.T.C. Stoof,$^1$ and 
        E.A. Cornell$^2$}
\address{$^1$Institute for Theoretical Physics,
             University of Utrecht, Princetonplein 5, \\
             P.O. Box 80.006, 3508 TA  Utrecht,
             The Netherlands \\
         $^2$JILA, National Institute of Standards and Technology, 
             University of Colorado \\
             and\\
             Physics Department, University of Colorado, Boulder CO 
             80309-0440} 
\maketitle

\begin{abstract}
We investigate the possibilities of distinguishing the mean-field and
fluctuation effects on the critical temperature of 
a trapped Bose gas with repulsive interatomic interactions. 
Since in a direct measurement of the critical temperature as a function 
of the number of trapped atoms these effects are small
compared to the ideal gas results, we propose to 
observe Bose-Einstein condensation by adiabatically ramping 
down the trapping frequency. Moreover, analyzing this adiabatic cooling 
scheme, we show that fluctuation effects can lead to
the formation of a Bose condensate at frequencies which are 
much larger than those predicted by the mean-field theory. 
\end{abstract}


\begin{multicols}{2}
The experimental observation of Bose-Einstein condensation in atomic 
gases of $^{87}$Rb \cite{eric}, $^{23}$Na \cite{wolfgang}, and 
$^7$Li \cite{randy} confined in a magnetic trap, 
has enormously boosted the theoretical and experimental research of such 
quantum gases. After studying the elementary excitations of the
condensate \cite{cornell,ketterle}, the JILA and MIT groups have also
started to determine more accurately the critical temperature of the 
gas as a function of the number of atoms in the trap, and in particular 
the deviation of this temperature from the ideal case \cite{jim}. 
Measuring these deviations will give important information on the effect 
of interactions on the thermodynamic properties of the gas and may also
signal a break-down of the Bogoliubov theory that has been so succesful
in explaining the collective excitation spectrum 
\cite{fetter,sandro,singh,keith}. 
It is therefore of considerable theoretical interest.

For a non-interacting gas in an isotropic external harmonic potential
$V({\bf r}) = m \omega^2 {\bf r}^2/2$, it is well known that if the number of 
particles $N \gg 100$, Bose-Einstein condensation occurs at a temperature 
\begin{equation}
\label{tempkritiek0}
T_0 = \left( \frac{N}{\zeta(3)} \right)^{1/3} \frac{\hbar \omega}{k_B}~.
\end{equation}
However, interactions between the atoms influence the critical behavior of 
the alkali gases of interest and shift the critical temperature. Roughly
speaking there are two effects to consider. First, the inhomogeneity
of the gas leads (above the transition) to a spatially-varying mean-field 
potential of $2n({\bf r})T^{2B}$, where $n({\bf r})$ is the 
local density of the gas, $T^{2B} = 4\pi a\hbar^2/m$ is the two-body 
transition matrix at zero energy and momenta, and $a$\ is the 
$s$-wave scattering length. Depending on the sign of the scattering length
the atoms are therefore either repelled from ($a>0$) or attracted to ($a<0$)
the center of the trap and it requires more or less atoms respectively, to 
acquire the necessary central density for Bose-Einstein condensation. Using a 
local-density approximation this effect was recently studied by Giorgini,
Pitaevskii and Stringari \cite{stringari}. They obtained a 
mean-field shift in the critical temperature of
\begin{equation}
\label{tempkritiekint}
\frac{T_c - T_0}{T_0} \simeq -1.33 \frac{a}{l} N^{1/6},
\end{equation}
where $l=\sqrt{\hbar/m\omega}$ is the extent of the harmonic oscillator 
ground state. Secondly, Bijlsma and Stoof have found by means of a 
renormalization group calculation that for gases with a positive scattering 
length the necessary central density for Bose-Einstein condensation 
is reduced considerably due to fluctuation effects \cite{michel}. This means 
that the degeneracy parameter $n({\bf 0})\Lambda_{th}^3$ (with 
$\Lambda_{th} = (2\pi\hbar^2/m k_B T)^{1/2}$ the thermal de Broglie wavelength)
at the critical temperature is not $\zeta(3/2) \simeq 2.612$, as assumed by 
the mean-field theory, but instead a monotonically decreasing function
of the quantum parameter $a/\Lambda_{th}$. It is therefore always smaller than
the ideal gas value $\zeta(3/2)$. 

Although the latter effect clearly increases the critical temperature, it is
at present unclear to what amount this increase cancels the
mean-field shift in the critical temperature. To investigate this issue,
we need quantitative information on the density profile of 
a trapped Bose gas with repulsive interactions, which we obtain by
performing numerical calculations in the local-density approximation. 
In this approximation the 
system is considered to be locally homogeneous, and the inhomogeneity 
induced by the trapping potential $V({\bf r})$ and the mean-field interaction
$2n({\bf r}) T^{2B}$ is determined selfconsistently by the spatially-dependent 
effective chemical potential 
$\mu'({\bf r}) = \mu - V({\bf r}) - 2 n({\bf r}) T^{2B}$, where $\mu$ is the
actual chemical potential of the gas. 

For our purposes, which will become clear shortly, we also need to calculate 
the total number of particles $N$ in the gas, together with the entropy $S$. 
According to statistical physics, these quantities can be calculated from
the grand-canonical potential $\Omega$ of the system, i.e. from
\begin{eqnarray}
N = - \left. \frac{\partial \Omega}{\partial \mu} \right|_{T,V} ~{\rm and}~ 
S = -\left. \frac{\partial \Omega}{\partial T} \right|_{\mu,V}. \nonumber
\end{eqnarray}
Above the critical temperature the (mean-field) grand-canonical potential 
of an interacting Bose gas is given by 
\begin{eqnarray}
\label{omega}
\Omega = k_BT \sum_j \ln(1 - \exp[&-&\beta(\epsilon_j - \mu)]) \nonumber \\
       & - & \int d{\bf r}\ n^2({\bf r})T^{2B}~, 
\end{eqnarray}
where $\epsilon_j$ denote the one-particle energy levels 
in the effective potential $V({\bf r}) + 2n({\bf r})T^{2B}$ \cite{marianne}.
Implementing the local-density approximation by putting 
$\epsilon_j = \hbar^2 {\bf k}^2/2m + V({\bf r}) + 2n({\bf r})T^{2B}$ and
using $\sum_j = \int d{\bf r} \int d{\bf k}/(2\pi)^3$, we can perform the 
gaussian integral over the momenta $\hbar{\bf k}$ to find   
\begin{eqnarray}
 - \frac{k_B T}{\Lambda_{th}^3} \int d{\bf r}\ g_{5/2}[z({\bf r})] \nonumber
\end{eqnarray}
for the first term on the right-hand side of Eq.~(\ref{omega}). Here we
introduced the Bose function   
\begin{equation}
\label{g52}
g_{5/2}(z) = -\frac{4}{\sqrt{\pi}} \int_0^{\infty} dx~ x^2
              \ln{(1 - z \exp{[-x^2]})}
\end{equation}
and the local fugacity $z({\bf r}) = \exp{[\beta \mu'({\bf r})]}$.

Calculating now the derivative of $\Omega$ with respect to $\mu$, 
one obtains for the total number of particles
\begin{equation}
\label{partnumber}
N = \int d{\bf r}~ n({\bf r}) 
   = \frac{1}{\Lambda_{th}^3} \int d{\bf r}~ g_{3/2}[z({\bf r})],
\end{equation}
where we used $g_{3/2}(z) = z [d g_{5/2}(z) / d z]$. Performing a
similar differentiation with respect to the temperature, the entropy
of the system is found to be
\begin{equation}
\label{entropie}
S= k_B \int d{\bf r}~ \left\{ \frac{5}{2} g_{5/2}[z({\bf r})] 
                           - n({\bf r}) \ln[z({\bf r})] \right\}.
\end{equation}
Note that in both differentiations we used the fact that $z({\bf r})$ has an 
implicit dependence on the chemical potential and the temperature through the 
density $n({\bf r})$. The contributions to the particle number and the entropy 
from this dependence exactly cancel, however,
the contributions coming from the second term in the right-hand side of
Eq.~(\ref{omega}).  

\begin{figure}[htbp]
\psfig{figure=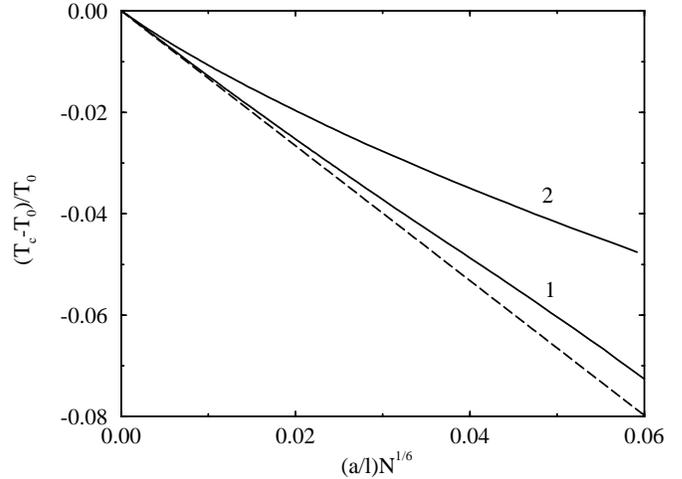}
\caption{\narrowtext
         Relative change in the critical temperature for an interacting gas 
         in a harmonic oscillator trap. Curve 1 gives the  
         mean-field result and curve 2 includes also the effect of
         fluctuations. The dashed line represents Eq.~(2).
         \label{fig1}}
\end{figure}

At this point we are ready to discuss the way in which we numerically 
determine the critical temperature of the gas as a function of
the number of particles in the trap and, in particular, how we include 
the fluctuation effects on this critical temperature.
To include these fluctuations in a first approximation, we calculate
at a given temperature the value of the quantum parameter $a/\Lambda_{th}$ and
extract from the appropriate value of the critical degeneracy parameter
\cite{michel} the value of the central density $n({\bf 0})$ at which 
Bose-Einstein condensation first occurs. Knowing the critical central
density we can also immediately determine the chemical potential at 
criticality from $n({\bf 0}) \Lambda_{th}^{3} = g_{3/2}[\exp{\{\beta
(\mu - 2 n({\bf 0}) T^{2B})\}}]$. 
The corresponding number of particles is then self-consistently 
calculated from Eq.~(\ref{partnumber}). To compare with the
mean-field results, we also calculate the number of particles in the
system for a central density of $\zeta(3/2)/\Lambda_{th}^3$. Both results, 
together with Eq.~(\ref{tempkritiekint}) are plotted in Fig.~\ref{fig1}. 

For different values of the scattering length $a$ and the
trapping frequency $\omega$, we determined the critical temperature 
as a function of the number of particles and found that for both
calculations the relative shift in the critical temperature is a 
universal function of $(a/l)N^{1/6}$ as indicated in Fig.~\ref{fig1}. 
From this figure we see that Eq.~(\ref{tempkritiekint}) gives the 
correct lowest order mean-field shift of the temperature, but also that
higher order corrections become important relatively soon. Moreover, the
fluctuation effects are roughly speaking of the same order of magnitude 
under the experimental conditions of interest, although they are 
clearly unable to overcome the decrease of the critical temperature due
to the mean-field potential. However, at this point we have to remind
ourselves that the present calculation underestimates the 
influence of fluctuations because it does not incorporate the effect that
fluctuations have on the density profile in the trap. In principle, this
could also be included by performing what one might call a `local 
renormalization group' calculation, i.e.\ apply the renormalization group 
theory of Ref.~\cite{michel} at each point in space.  

Since the critical temperature of an interacting Bose gas in an 
absolute sense does not deviate much from the ideal gas value, we now
propose a possible experiment that could make the effects of the 
interatomic interactions visible
without having to deal with a large ideal gas `background'.
Suppose we have a gas of $N$ particles in an
isotropic trap with frequency $\omega$ and 
at a temperature $T$\ above the critical temperature. 
The gas also has a certain entropy given by Eq.~(\ref{entropie}). 
We now ask ourselves what would happen if the system 
is cooled adiabatically to a lower temperature by slowly lowering the 
trapping frequency to a value $\omega'$. Since during this
adiabatic process the entropy and the number of particles in the gas remain 
constant, we know the answer to this question if we determine the final 
chemical potential and temperature 
combination $(\mu',T')$ such that $N(\mu',T',\omega')=N(\mu,T,\omega)$ and
$S(\mu',T',\omega')=S(\mu,T,\omega)$.

The reason why we propose to adiabatically cool the gas in this manner is that
for the ideal Bose gas, the only way to keep both
the entropy as well as the particle number constant during the change 
$\omega \rightarrow \omega'$ in the trapping frequency, is to change the 
temperature from $T$ to $(\omega'/\omega)T$ and  
the chemical potential from $\mu$ to $(\omega'/\omega)\mu$. This leaves all the
occupation numbers of the harmonic oscillator states unchanged. More
importantly, it leaves the fugacity and therefore the degeneracy parameter 
$n({\bf 0})\Lambda_{th}^3$ in the center of the trap unchanged. 
So for the ideal gas, 
adiabatic cooling does not bring us closer to criticality. Therefore, if
we can reach Bose-Einstein condensation in an interacting gas in this way, 
it must be entirely due to the effect of interactions.

\begin{figure}[htbp]
\psfig{figure=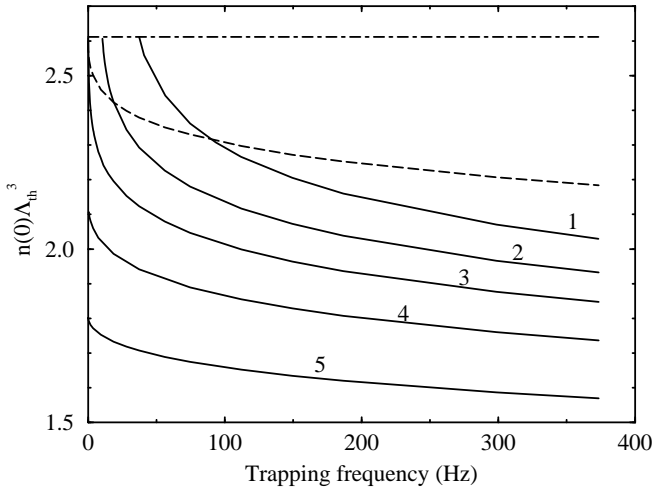}
\caption{\narrowtext
         Adiabatic trajectories (solid lines) of a gas of 
         $20,000$\ $^{87}$Rb particles when the trapping frequency is ramped 
         down.
         The initial temperatures are 1) 442.21 nK, 2) 445.85 nK, 3) 449.43 nK,
         4) 454.72 nK, and 5) 464.16 nK. 
         The dashed line indicates the value for the central degeneracy
         parameter 
         at which Bose-Einstein condensation first occurs if fluctuation 
         effects are taken into account. The dot-dashed line 
         denotes the mean-field critical degeneracy parameter. 
         \label{fig2}}
\end{figure}

In Fig.~\ref{fig2}, the value of the degeneracy parameter in the center of 
the trap 
is plotted for five different adiabatic processes (solid lines 1 to 5) during 
an adiabatic ramp down of the trap frequency from a starting frequency of 
$373$\ Hz. All five adiabatic curves correspond to a gas containing $N=20,000$\ 
$^{87}$Rb particles with an $s$-wave scattering length of $109 a_0$.
The initial temperature of the gas increases from the
upper to the lower curves and consequently the chemical potential increases 
from the lower to the upper curve, since the total number of particles is
the same. Physically, for a nonideal gas the extra term 
$2 n({\bf r}) T^{2B}/k_BT$ in the exponent of $z({\bf r})$ causes a
deviation from a horizontal line, because the density does not 
scale linearly with $T$ but approximately with $T^{3/2}$. Thus keeping
$N$ and $S$ constant when $\omega$ is varied, we now have to change the local
fugacity and as a result the degeneracy parameter in the center of the
trap also changes. Furthermore, to achieve Bose-Einstein condensation
in a gas with a positive scattering length, it is clear that the frequency
must be ramped down because this lowers the density of the gas
and reduces the effect of the repulsive interaction $2n({\bf r})T^{2B}$. 
As a result, the density in the center of the trap reduces less than in the 
ideal gas case and the degeneracy parameter $n({\bf 0})\Lambda_{th}^3$\ 
increases.   
 
In Fig.~\ref{fig2} we also plotted the value of the 
critical degeneracy parameter 
following from the renormalization group theory of Ref.~\cite{michel} and
corresponding to the instantaneous temperature during the adiabatic cooling 
of the gas.  It should be noted that in principle this curve is different 
for each adiabatic 
trajectory. However, for the five trajectories presented here the 
differences cannot be distinguished on this scale. Another aspect that 
can be seen from Fig.~\ref{fig2} is that only the upper three
curves reach the conditions for Bose-Einstein condensation,
whereas curves 4 and 5 remain uncondensed all the way down to $\omega=0$ 
which is equivalent to $T=0$. Indeed, a detailed
analysis of the chemical potential and the central density
as a function of the instantaneous temperature 
$T$ shows that for curves 1 to 3 $\mu \propto T^{\alpha}$ and 
$n({\bf 0}) \propto T^{\beta}$ 
for the complete $\omega$-interval considered. Moreover, the powers obey
$\alpha < \beta <3/2$ which implies that although initially 
the gas is not condensed and $\mu< 2n({\bf 0}) T^{2B}$, there 
always will be some lower and nonzero
temperature such that $\mu = 2 n({\bf 0}) T^{2B}$ and the gas fulfills
the mean-field critical conditions. Consequently it will have already 
achieved the critical conditions including fluctuation effects at a higher 
temperature.

However, for curves 4 and 5 the chemical potential and central density
exhibit a different behavior. This is due to
the fact that during the cooling process the chemical potential
becomes negative in order to satisfy conservation of both
the number of particles $N$ as well as the entropy $S$. 
Therefore, the system cannot become Bose condensed unless 
$\mu \uparrow 0$ and $2n({\bf 0})T^{2B} \downarrow 0$ for $T \rightarrow 0$.  
Looking again in more detail to the power behaviour of $\mu$ and $n({\bf 0})$ 
as a function of $T$, we find that $n({\bf 0}) \propto T^{\beta}$ with
$\beta$ almost equal to 3/2, and that $\mu \propto -T$ for $T \rightarrow 0$.
So both $\mu$ and $2n({\bf 0})T^{2B}$
indeed approach zero, but the limit of the ratio $\mu/k_BT$ will 
now decide whether or not the system achieves Bose-Einstein condensation
at zero temperature and frequency.  

\begin{figure}[htbp]
\psfig{figure=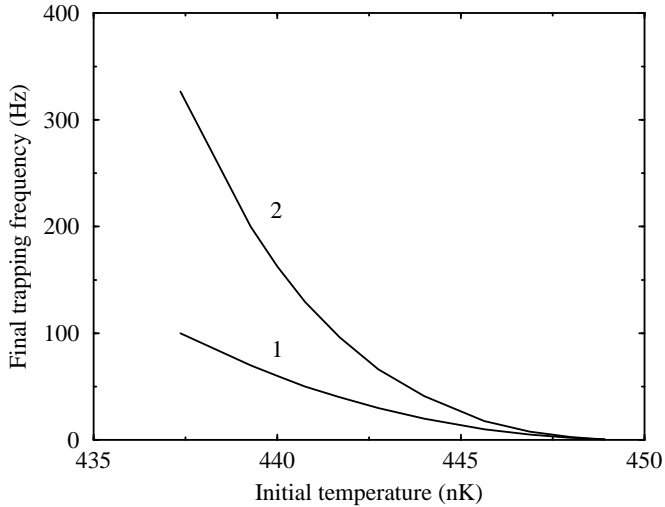}
\caption{\narrowtext
         Trapping frequency needed to obtain the critical conditions 
         in the center of the trap, starting from a configuration of 
         20,000 particles in a $373$\ Hz trap and at a temperature 
         denoted by the ordinate. Curve 1 is the mean-field result and
         curve 2 includes also the effect of fluctuations. 
         \label{fig3}}
\end{figure}

Fig.~\ref{fig3} shows a plot of the trapping frequencies
at which Bose-Einstein condensation first occurs if only mean-field effects
are taken into account (curve 1) and if also fluctuation effects are included
(curve 2). Again all adiabatic processes are for a gas containing
$20,000$ particles and an initial trapping frequency 
of $373$\ Hz. The initial temperature of the gas is given by the ordinate. 
From this figure it is clear that the corrections due to fluctuations
have a large influence on the final frequency at which Bose-Einstein 
condensation is achieved. For suitable initial conditions, 
the final trapping frequency can exceed the mean-field prediction by  
more than a factor of 3. We therefore conclude that in principle 
the fluctuation corrections are experimentally observable by 
adiabatically cooling the gas.

However, it must be noted that while curves 1 and 2 in Fig.\ 3 can be
separated by as much as a factor of three in 
the frequency axis, they are separated by only 1\% on the temperature axis.
This should be compared with the JILA group's most
recent attempt at absolute thermometry, with claimed {\it absolute} accuracy 
no better than 5\% \cite{jim}. Nevertheless, measurements of the condensate 
fraction near the critical temperature
indicate that the shot-to-shot {\it reproducibility}
of the scaled temperature is much better, i.e.\ on the order of 
1\% \cite{deborah}. 
Moreover, the scaled temperature varies smoothly with final
evaporative rf cut, which is a readily adjustable experimental 
parameter.
We therefore envision an experiment in which samples are prepared
at temperatures precisely determined relative to
the empirical onset temperature of BEC, and then subjected to adiabatic ramps
of the trapping frequency. 

In summary, we showed that, although fluctuation corrections  
decrease the critical density of a trapped gas of repulsively interacting 
bosonic atoms, they do not completely cancel the increase in 
the critical number of particles obtained from mean-field theory.  
The effect of the interactions, and in particular, of the fluctuation
corrections to the mean-field theory can be observed more directly by 
adiabatically cooling the gas to criticality. Here we imply changing 
adiabatically the frequency of a harmonic trap and not the power law of the 
trapping potential, which has recently been shown to change already 
the degeneracy parameter of the ideal Bose gas \cite{jook}. In addition to
being more difficult to implement experimentally, the latter approach
is not so favorable for our purposes because it is less sensitive to the 
effect of interactions. We hope that this work might 
stimulate an experimental observation of these interesting phenomena in the 
near future.

We acknowledge stimulating conversations with Michel Bijlsma, Jason Ensher,
Deborah Jin, and Sandro Stringari. E.C. acknowledges support from the 
O.N.R. and the N.S.F.

\end{multicols}
\end{document}